\documentclass{article}
\usepackage{spconf,amsmath,graphicx,hyperref, multirow, booktabs, array}
\usepackage{enumitem}
\usepackage{makecell}
\usepackage[table]{xcolor}

\title{PHONOLOGICAL TOKENIZER: PROSODY-AWARE PHONETIC TOKEN \\
VIA MULTI-OBJECTIVE FINE-TUNING WITH DIFFERENTIABLE K-MEANS}

\name{Kentaro Onda$^{\star \dagger}$ \quad 
Hayato Futami$^{\dagger}$ \quad
Yosuke Kashiwagi$^{\dagger}$ \quad 
Emiru Tsunoo$^{\dagger}$ \quad
Shinji Watanabe $^{\ddagger}$}
  
\address{$^{\star}$ The University of Tokyo \qquad
      $^{\dagger}$Sony Group Corporation \qquad
      $^{\ddagger}$Carnegie Mellon University}
\begin{document}
\ninept
\maketitle
\begin{abstract}
In recent years, there has been growing interest in representing speech with discrete tokens, which serve as pseudo-text for speech language models (speechLMs) and as efficient intermediate representations for downstream tasks. These tokens are typically categorized as acoustic and phonetic tokens: the former holds detailed acoustic information for reconstruction while the latter mainly captures linguistic content.
In human speech communication, however, unnecessary acoustic details such as speaker information are abstracted, while both linguistic and prosodic information are utilized for speech comprehension and production. Given this, neither type of token seems an ideal representation for tasks sensitive to prosody, such as speechLMs. In this study, we propose the \textit{Phonological Tokenizer}, a method that fine-tunes phonetic tokens via differentiable k-means with a multi-task objective of ASR and speech resynthesis.
Experimental validation on diverse tasks confirms that our tokens retain phonological (both linguistic and prosodic) information while appropriately discarding speaker identity.
\end{abstract}
\section{Introduction}
\label{sec:intro}
In recent years, there has been increasing research interest in representing and processing speech as a sequence of discrete tokens \cite{guo2025recentadvancesdiscretespeech, mousavi2025discreteaudiotokenssurvey}. These discrete tokens are utilized as “pseudo-text” in speech language models (speechLMs) \cite{lakhotiaetal2021generative, arora2025landscapespokenlanguagemodels} and serve as efficient intermediate representations for various tasks such as automatic speech recognition (ASR) and text-to-speech (TTS) \cite{mousavi24_interspeech, chang24b_interspeech}. Generally, discrete tokens can be categorized into two types: acoustic tokens and phonetic tokens \footnote{Phonetic tokens are sometimes also referred to as \textit{semantic} tokens, but their actual property is closer to phoneme-like units \cite{wells22_interspeech,shi2023, choi24b_interspeech}}.
Acoustic tokens are typically learned using VQ-VAE-based models with the goal of reconstructing speech waveforms \cite{zeghidour2021soundstream, fossez2023high}. They retain detailed acoustic information, including speaker identity and background noise, and are considered suitable for speech synthesis tasks. On the other hand, phonetic tokens are obtained by applying k-means clustering to the outputs of pre-trained self-supervised learning (SSL) models \cite{hubert}, and are regarded as being more suitable for extracting linguistic information \cite{yang2024towards}.

However, considering how human speech communication functions, both types of tokens appear somewhat extreme. Humans abstract away unnecessary acoustic details such as voice timbre or background noise when perceiving speech, while combining prosody with linguistic information for both speech comprehension and production \cite{cutler1997prosody, nespor2007prosodic}. Therefore, rather than retaining all the acoustic details or reducing speech to purely linguistic content, what is desirable is a representation that captures the holistic \textit{phonological} aspects of speech, that is, linguistic content and prosody. 
This type of token will have intermediate properties between acoustic and phonetic tokens.
Several recent studies partially address this need by incorporating a pretrained SSL model into the learning of acoustic tokens to enhance their ability to capture linguistic information \cite{zhangspeechtokenizer, defossez2024moshispeechtextfoundationmodel, ye2025codec}, known as hybrid tokens. 
However, these methods are mainly based on the residual vector quantization (RVQ) framework and the overall representation remains essentially a multi-codebook acoustic token. Thus, fully leveraging the advantages of these tokens requires additional, somewhat complex architectures in downstream models to manage multiple streams effectively. Also, the use of multiple codebooks inherently reduces data compression efficiency, which is the key advantages of discrete representations \cite{chang2024exploring, wang-etal-2025-speech}.

In this study, we propose the \textit{Phonological Tokenizer}: a single-codebook speech tokenizer that captures the holistic phonological aspects of speech, namely linguistic and prosodic information, while discarding unnecessary acoustic details such as background noise and speaker identity. We leverage the flexible discretization capability of the recently proposed differentiable k-means \cite{onda2025differentiablekmeansfullyoptimizeddiscrete} to build this tokenizer. We fine-tune phonetic tokens obtained from a pre-trained SSL model using differentiable k-means in a multi-objective framework combining ASR and speech resynthesis.
The resulting tokenizer demonstrates high performance in both speech understanding and generation tasks, as well as in speechLM applications.

The key strengths of our approach are summarized as follows:
\begin{itemize}[left=0pt]
    \item \textbf{Balancing prosody preservation and speaker information removal:} We fine-tune phonetic tokens for both ASR and speech resynthesis with weighted losses, along with conditioning the vocoder on speaker embeddings during training. 
    As a result, our method effectively incorporates prosodic information while preserving the ability of phonetic tokens to capture linguistic information and discard speaker information. 
    It exhibits especially strong performance in tasks where prosody is crucial, such as emotion recognition, voice conversion, and speechLMs.


    \item \textbf{High compression efficiency:} By optimizing SSL-based phonetic tokens using differentiable k-means \cite{ onda2025differentiablekmeansfullyoptimizeddiscrete, Gao2020}, our method enables fine-tuning of token properties while maintaining a single codebook. This achieves significantly higher data compression efficiency compared to multi-codebook acoustic tokens, while showing superior or comparable performance to baseline tokenizers \cite{zhangspeechtokenizer, Chen2021WavLMLS, jiwavtokenizer} on many tasks.

    \item \textbf{Reduced training data requirement:} Since our approach fine-tunes a pre-trained large-scale speech foundation model (WavLM-large \cite{Chen2021WavLMLS}), it can build a versatile speech tokenizer with only small training data. In this study, we fine-tune phonetic tokens using 44 hours of additional training data (VCTK corpus \cite{veaux2017cstr}) to modify its properties.
    This represents a substantially smaller data requirement compared to prior studies \cite{zhangspeechtokenizer, jiwavtokenizer} using large-scale datasets such as LibriSpeech (960h) \cite{libri} or LibriTTS (585h) \cite{zen19_interspeech}.
\end{itemize}

\section{Related Works}

\subsection{Hybrid tokens utilizing pretrained SSL models}
Several prior studies have proposed hybrid tokens that extend RVQ-based acoustic tokens by integrating pre-trained SSL models to better capture linguistic content \cite{zhangspeechtokenizer, defossez2024moshispeechtextfoundationmodel, ye2025codec}. These tokens demonstrate superior language understanding capabilities compared to acoustic tokens, which are trained solely for reconstruction.
However, as discussed in Introduction, these tokens consist of multiple codebooks, which results in low efficiency and limited usability.
In this study, we propose hybrid tokens based on phonetic tokens by fine-tuning them using differentiable k-means, enabling a single codebook. 

\subsection{Supervised tokenizers}
Several studies have proposed ASR-based tokenizers that allow tokens to better capture linguistic information \cite{rubenstein2023audiopalm,du2024cosyvoice}.
Our previous work \cite{onda2025differentiablekmeansfullyoptimizeddiscrete}, which optimized phonetic tokens for ASR using differentiable k-means, can also be categorized as this type of tokens. In this work, we further extend this idea by fine-tuning phonetic tokens with multi-objective of ASR and speech resynthesis.

\subsection{Disentanglement-oriented codec tokens}
Several prior studies have proposed codec-based tokens designed to promote disentanglement by representing global information, such as speaker identity, in separate branches \cite{ren2024fewer,facodec, guo25_interspeech}. However, approaches based on phonetic tokens have not yet been explored. In this work, we aim to incorporate prosodic information into phonetic tokens in which speaker information is already suppressed, through fine-tuning within a multi-task learning framework.

\section[Multi-objective Optimization with Differentiable K-means]{Multi-objective Optimization\\ with Differentiable K-means}
\begin{figure}[t]
  \centering
  \includegraphics[width=0.95\linewidth]{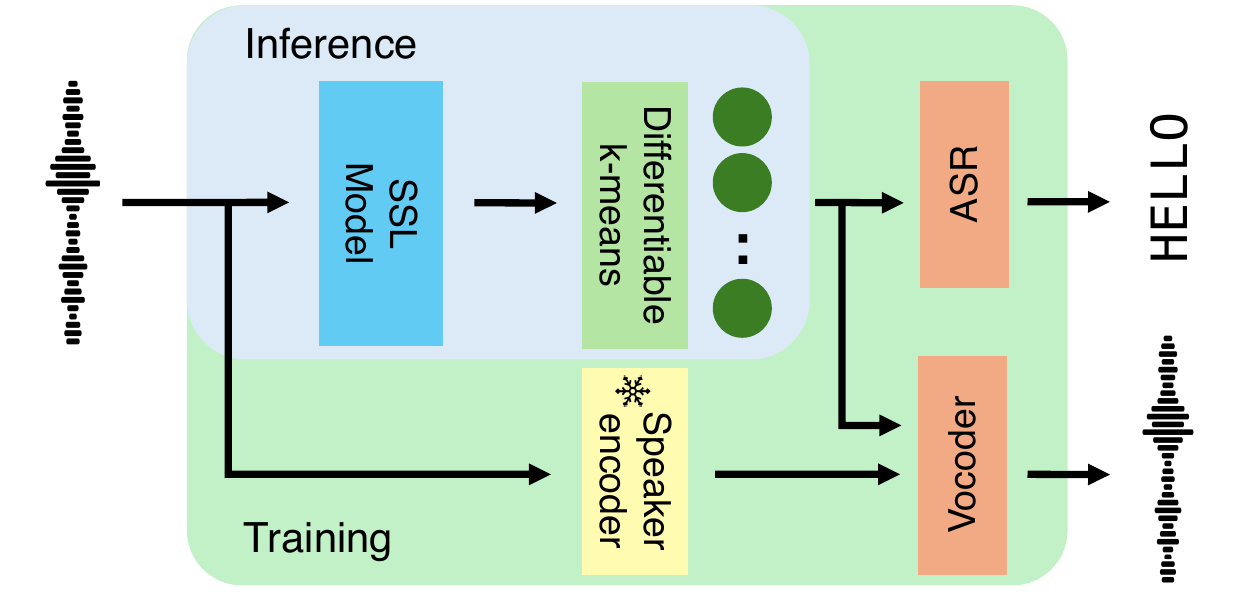}
  \vspace*{-2mm}
  \caption{Architecture of the Phonological Tokenizer: multi-objective fine-tuning of SSL-derived phonetic tokens}
  \label{fig:model}
  \vspace*{-4mm}
\end{figure}
\subsection{Phonetic token optimization via differentiable k-means}
 Our previous study \cite{onda2025differentiablekmeansfullyoptimizeddiscrete} proposed to optimize discrete tokens obtained from SSL models for specific purposes by introducing differentiable k-means. The ASR loss $\mathcal{L}^{\text{asr}}$ is used to jointly optimize all components of the model: 1) the SSL model $\theta_{\text{ssl}}$, performing feature extraction $\mathrm{SSL}(X;\theta_{\text{ssl}})$ from the input speech $X$; 2) the cluster centroids $M$, used for the differentiable k-means $\mathrm{DiffKM}(\cdot; M)$ to discretize the SSL features; and 3) the ASR model $\theta_{\text{asr}}$, predicting the text transcription $Y$ from the token sequence via $\mathrm{ASR}(\cdot;\theta_{\text{asr}})$.
{\setlength{\abovedisplayskip}{4pt}
\setlength{\belowdisplayskip}{4pt}
\setlength{\abovedisplayshortskip}{4pt}
\setlength{\belowdisplayshortskip}{4pt}
\begin{align}
\label{eq:joint_loss}
\mathcal{L}^{\text{asr}}\Bigl(
    Y,\,
    \mathrm{ASR}\bigl(
        \mathrm{DiffKM}\bigl(\mathrm{SSL}(X;\,\theta_{\text{ssl}});\,M\bigr);\,
        \theta_{\text{asr}}
    \bigr)
\Bigr)
\end{align}
}
This approach not only improved the accuracy of ASR but also modified the properties of the tokens, enabling them to represent purer linguistic information. This demonstrates the effectiveness of differentiable k-means in manipulating the properties of discrete tokens.

\subsection{Proposed method: multi-objective optimization}
In this study, we propose to extend the loss function in Eq. (\ref{eq:joint_loss}) by incorporating a weighted reconstruction loss, to achieve \textit{Phonological Tokenizer}, which retains linguistic and prosodic information while appropriately removing speaker information. 
The ASR loss $\mathcal{L}^{\text{asr}}$ encourages the extraction of linguistic information while suppressing prosody and speaker information. In contrast, the reconstruction loss $\mathcal{L}^{\text{voc}}$ drives the tokens to capture all the acoustic details, including prosody and speaker identity.
Thus, these two losses can be seen as tuning the token properties toward those of the phonetic and acoustic tokens, respectively.
Therefore, to obtain the Phonological Tokenizer that has intermediate properties between phonetic and acoustic tokens, it is reasonable to balance these two losses and optimize the tokens in a multi-task manner.
{\setlength{\abovedisplayskip}{4pt}
\setlength{\belowdisplayskip}{4pt}
\setlength{\abovedisplayshortskip}{4pt}
\setlength{\belowdisplayshortskip}{4pt}
\begin{align}
\label{eq:joint_loss_voc}
&\mathcal{L}
=
(1-\alpha)\, \mathcal{L}^{\text{asr}}\Bigl(
    Y,\,
    \mathrm{ASR}\bigl(
        \mathrm{DiffKM}\bigl(\mathrm{SSL}(X;\,\theta_{\text{ssl}});\,M\bigr);\,
        \theta_{\text{asr}}
    \bigr)
\Bigr)
\nonumber\\[-2mm] 
&+ \alpha\, \mathcal{L}^{\text{voc}} \Bigl(
    X,\,
    \mathrm{Voc}\bigl(
        \mathrm{DiffKM}\bigl(\mathrm{SSL}(X;\,\theta_{\text{ssl}});\,M\bigr), E_{\text{spk}};\,
        \theta_{\text{voc}}
    \bigr)
\Bigr)
\end{align}
}
We weight these two losses
using $\alpha$. Both the ASR model $\theta_{\text{asr}}$ used for transcription $\mathrm{ASR}(\cdot;\theta_{\text{asr}})$ and the vocoder $\theta_{\text{voc}}$ for resynthesis $\mathrm{Voc}(\cdot;\theta_{\text{voc}})$ are trained using the discrete tokens obtained through differentiable k-means as a shared input. Thus, 
the tokens
are optimized for both tasks in a balanced way. In order to help disentangle speaker information from the tokens, we employed a pre-trained speaker encoder to provide the speaker embedding $E_{\text{spk}}$ as an auxiliary conditioning input to the vocoder.

An overview of our tokenizer is presented in Fig. \ref{fig:model}.
During training, the entire module, except for the speaker encoder, is jointly optimized. At inference time, discrete tokens are generated using only the fine-tuned SSL model $\theta_{\text{ssl}}$ followed by differentiable k-means with learned cluster centroids $M$.

\section{Experiments}

\subsection{Experimental setup}
The model training was conducted using ESPnet \cite{watanabe18_interspeech}. The configuration related to differentiable k-means followed the settings described in \cite{onda2025differentiablekmeansfullyoptimizeddiscrete}, and the cluster size was set to 2000. For the SSL model, we used the 21st layer of WavLM-large  \cite{Chen2021WavLMLS},
and the cluster centroids $M$ were initialized using standard k-means clustering, trained on a 30-hour subset of the LibriSpeech-100h dataset \cite{libri}. For ASR, we employed the joint CTC/attantion-based encoder-decoder (AED) model \cite{ctcaed}, and for the vocoder, we used HiFi-GAN \cite{kong2020hifi}\footnote{\url{https://github.com/kan-bayashi/ParallelWaveGAN}} . As the speaker encoder, we adopted a pretrained ECAPA-TDNN model \cite{DBLP:conf/interspeech/DesplanquesTD20}\footnote{\url{https://hf.co/speechbrain/spkrec-ecapa-voxceleb}}.
As described in \cite{onda2025differentiablekmeansfullyoptimizeddiscrete}, training was conducted in two stages. In the first stage, the SSL model $\theta_{\text{ssl}}$ and cluster centroids $M$ were kept frozen, and only the ASR $\theta_{\text{asr}}$ and vocoder $\theta_{\text{voc}}$ components were trained for 30 epochs with a learning rate of 1e-4.
In the second stage, the entire module, including the SSL model $\theta_{\text{ssl}}$ and centroids $M$ (except the speaker encoder) was fine-tuned for 60 epochs with a learning rate of 1e-5.
The vocoder training included adversarial learning, with the discriminator updated concurrently throughout both stages.

Training was conducted using the VCTK corpus \cite{veaux2017cstr} with speed perturbation ($\times$0.9, 1.0, and 1.1). We adopted $\alpha=0.1$\footnote{See Sec.\ref{subsec: ablation} for the results of the ablation study} as the weight for the vocoder loss (in Eq.(\ref{eq:joint_loss_voc})).
For reference, we also show the results for $\alpha=0$ and $\alpha=1$, where the tokens are fine-tuned with single-task objectives of ASR and vocoder, respectively.
In the following experiments, we used the tokens obtained from the trained model to perform various discrete token-based speech tasks, and compared their performance with that of existing models. As baselines, we used WavLM (21st layer) followed by standard offline k-means clustering (k=2000) as phonetic token.
We also used SpeechTokenizer \cite{zhangspeechtokenizer}
(using only the first codebook) trained on LibriSpeech as hybrid token, and WavTokenizer \cite{jiwavtokenizer}
trained on LibriTTS as acoustic token. We present a comparison of the basic properties of these baseline tokens and our proposed tokens in Table \ref{tab:bitrate}. This shows that once initialized with phonetic tokens, our tokens can be effectively fine-tuned with very limited additional data, while keeping high compression efficiency.

\begin{table}[tb]
	\centering
	\caption{Statistics of baseline and proposed tokens}
	\label{tab:bitrate}
    \resizebox{\columnwidth}{!}{
	\begin{tabular}{l@{\hspace{6pt}}l@{\hspace{3pt}}l|cccc}
	  \toprule
                 &            &  & Bit 
                 & Vocab. & Tokens & Train \\
                 &           &   & rate & size &/sec. & data \\\midrule
	\multirow{3}{*}{Baseline} 
         & Discrete WavLM &(phonetic)& 548.3 & 2000 & 50  & \phantom{1}30h\\
           & SpeechTokenizer &(hybrid)& 500.0 & 1024 & 50 & 960h \\
           & WavTokenizer& (acoustic)& 900.0 &  4096& 75 & 585h\\\midrule
        \textbf{Proposed}& \multicolumn{2}{@{\hspace{0pt}}l|}{\textbf{Phonological Tokenizer}} & 548.3 & 2000 & 50 &  30 + 44h \\   
                  
	  \bottomrule
	\end{tabular}
    }
    \vspace*{-5mm}
\end{table}

\subsection{Evaluation on discriminative tasks}
\label{subsec:discriminative}
\begin{table}[tb]
	\centering
	\caption{Discriminative task performance: ASR on librispeech-100 \& Emotion Recognition (ER) on Ravdess \& Speaker Identification (SID) on VoxCeleb. The best result in each column is \textbf{bolded}.}
	\label{tab:discriminative}
    \resizebox{\columnwidth}{!}{
	\begin{tabular}{l@{\hspace{6pt}}l@{\hspace{3pt}}l|c|c|c}
	  \toprule
                 &       &       & ASR 
                 & ER & SID \\
                 &       &       & WER (test-\{clean / other\}) & acc. & acc. \\
                 & & & ($\downarrow$) &  ($\uparrow$) & ($\uparrow$)\\\midrule
	\multirow{3}{*}{Baseline} 
         & Discrete WavLM & (phonetic) & \phantom{1}4.3/\phantom{1}7.1 & 41.7 & 27.7 \\
           &  SpeechTokenizer\phantom{1} & (hybrid) & \phantom{1}9.3/23.5 & 39.2 & 29.1  \\
           & WavTokenizer & (acoustic) & 96.7/96.8 &  24.2& \textbf{82.7} \\\midrule
        Single-task& ASR-only \cite{onda2025differentiablekmeansfullyoptimizeddiscrete}  &($\alpha = 0 $) & \textbf{\phantom{1}4.0/\phantom{1}7.0} & 41.7 & 20.6 \\   
        Optimized& Voc-only& ($\alpha = 1 $) & 10.4/27.7 & 40.0& 49.0\\\midrule  
         \multirow{2}{*}{\textbf{Proposed}}       &  \textbf{Phonological} & \multirow{2}{*}{($\alpha = 0.1 $)} & \multirow{2}{*}{\phantom{1}4.6/\phantom{1}8.5 }& \multirow{2}{*}{\textbf{51.7}} & \multirow{2}{*}{29.5} \\
                & \textbf{Tokenizer} & & &  & \\
	  \bottomrule
	\end{tabular}
    }
    \vspace*{-4mm}
\end{table}

We first evaluated the performance of Phonological Tokenizer on discriminative tasks. Focusing on three distinct aspects, linguistic information, prosody, and speaker identity, we trained downstream models for ASR, emotion recognition (ER), and speaker identification (SID). For ASR, we trained joint CTC/AED models using LibriSpeech-100h. For ER, we trained ECAPA-TDNN models on RAVDESS \cite{ravdess}, a dataset of the same sentences spoken with different emotions, 
using
a speaker-independent split.
For SID, we trained ECAPA-TDNN models on VoxCeleb1 \cite{nagrani2020voxceleb}.

The results are shown in Table \ref{tab:discriminative}. 

\noindent \textbf{ASR:} 
Our Phonological Tokenizer, while showing slightly lower performance than Discrete WavLM, demonstrated clear superiority over both SpeechTokenizer and WavTokenizer. This indicates that our token represents sufficient linguistic information. Considering the substantial drop observed when optimized solely for reconstruction ($\alpha=1$), our multi-task framework seems effective in preserving the capability of phonetic tokens for capturing linguistic content.

\noindent \textbf{ER:} The Phonological Tokenizer achieved by far the best performance.
This indicates that our proposed tokens successfully capture prosodic information in a speaker-independent manner. 


\noindent \textbf{SID:} The Phonological Tokenizer properly showed quite low accuracy as observed in Discrete WavLM and SpeechTokenizer, indicating minimal speaker information in the tokens.
The lower performance of Voc-only ($\alpha=1$) compared to WavTokenizer, despite being trained only for reconstruction, suggests that using an SSL model  and conditioning the vocoder with speaker embeddings are effective for disentangling speaker information from the tokens.

Overall, the results 
indicate that our Phonological Tokenizer successfully captures prosodic information (as shown in ER) while preserving the ability of phonetic tokens to capture linguistic information (in ASR) and suppress speaker information (in SID).

\subsection{Evaluation on generative tasks}
\label{subsec:generative}
\begin{table*}[tb]
	\centering
	\caption{Generative task performance: reconstruction on in-domain LJSpeech \& voice conversion on out-of-domain neutral read speech (TIMIT) and expressive speech (Expresso). The best result in each column is \textbf{bolded}, and those that outperform all baselines are \underline{underlined}.}
	\label{tab:generative}
    \resizebox{\textwidth}{!}{
	\begin{tabular}{l@{\hspace{6pt}}l@{\hspace{3pt}}l|cccc|cccc|cccc}
	  \toprule
                 &       &       & \multicolumn{4}{c|}{LJSpeech reconstruction (ID)}
                 & \multicolumn{4}{c|}{TIMIT VC (OOD)} & \multicolumn{4}{c}{Expresso VC (OOD)} \\
                 &     &         & MCD & F0 RMSE & UTMOS & WER & F0 corr. & SpkSim & UTMOS & WER & F0 corr. & SpkSim & UTMOS & WER  \\
                 & & & ($\downarrow$) & ($\downarrow$)& ($\uparrow$) & ($\downarrow$)& ($\uparrow$)& ($\uparrow$)& ($\uparrow$)& ($\downarrow$)& ($\uparrow$)& ($\uparrow$)& ($\uparrow$)& ($\downarrow$)\\\midrule
	\multirow{3}{*}{Baseline} 
         & Discrete WavLM& (phonetic) & 5.64 & 0.289 & 3.81 & 2.8 & 0.371 & 0.757 & 3.63 & 10.3  & 0.382& 0.737 & 3.47& \textbf{12.2}\\
           & SpeechTokenizer & (hybrid) & 5.35 & 0.270 & 3.91 & 3.3& 0.383& 0.726 & 3.53& 18.6& 0.388& 0.706& 3.13& 24.0\\
           & WavTokenizer & (acoustic) & 4.47 & \textbf{0.176} & \textbf{4.13} & \textbf{2.7} & 0.356 & 0.256 & 2.02 & 34.0 & 0.520& 0.352& 2.24 & 27.7\\\midrule
        Single-task & ASR-only \cite{onda2025differentiablekmeansfullyoptimizeddiscrete}  & ($\alpha = 0 $) & 5.77 & 0.300 & 3.82 & 2.9 & \underline{0.385}& 0.756& \underline{3.70}& 10.6& 0.391& \underline{\textbf{0.738}}& \underline{\textbf{3.61}}& 12.6\\   
        Optimized& Voc-only  & ($\alpha = 1 $) & \underline{\textbf{4.42}}& 0.183& 4.08& 3.3& \underline{\textbf{0.484}}& 0.695& \underline{3.70}& 16.4& \textbf{\underline{0.543}}&0.608 & 2.96& 26.8\\\midrule  
                \multirow{2}{*}{\textbf{Proposed}}&  \textbf{Phonological}   & \multirow{2}{*}{($\alpha = 0.1 $)} & \multirow{2}{*}{4.99} & \multirow{2}{*}{0.208}& \multirow{2}{*}{4.06}& \multirow{2}{*}{2.9}& \multirow{2}{*}{\underline{0.456}}& \multirow{2}{*}{\underline{\textbf{0.762}}}& \multirow{2}{*}{\underline{\textbf{3.88}}}& \multirow{2}{*}{\phantom{1}\textbf{\underline{9.8}}}& \multirow{2}{*}{\underline{0.538}}& \multirow{2}{*}{0.724}&\multirow{2}{*}{\underline{3.58}} & \multirow{2}{*}{12.6}\\ 
                & \textbf{Tokenizer} &&&&&&&&&&&&&\\
                   
	  \bottomrule
	\end{tabular}
    }
    \vspace*{-4mm}
\end{table*}
We then evaluated performances on generative tasks. We trained unit HiFi-GAN on the LJSpeech \cite{ljspeech} with the obtained tokens. 
The evaluation is done both on reconstruction on in-domain (ID) LJSpeech and on voice conversion (VC) on out-of-domain (OOD) corpora. In VC, the tokens from OOD speech are input to the LJSpeech-trained vocoder to generate speech in the voice of LJSpeech speaker.
If the tokens appropriately preserve only the linguistic and prosodic information, the output should maintain the spoken content and speaking style of the input while converting only the voice timbre to that of LJSpeech. As OOD data, we used TIMIT \cite{timit} as neutral read speech and Expresso \cite{nguyen23_interspeech} as expressive speech.
For ID reconstruction, we evaluated mel cepstral distortion (MCD) and F0 root mean square error (F0 RMSE). For OOD VC, we evaluated F0 correlation (F0 corr.) with the source speech and speaker similarity (SpkSim) with the target speaker of LJSpeech. To assess overall quality, we checked UTMOS \cite{saeki22c_interspeech} and WER computed from Whisper-large-v3 \cite{whisper} transcriptions for both tasks.

The results are shown in Table \ref{tab:generative}.

\noindent \textbf{LJSpeech reconstruction (ID):} 
Across all the metrics, our model outperformed or matched Discrete WavLM and SpeechTokenizer. Although not as good as WavTokenizer, the degradation in UTMOS and WER was minimal.
This demonstrates that our Phonological Tokenizer is sufficiently effective for generative tasks.
While our token does not retain fine-grained acoustic details sufficient for precise signal-level reconstruction, it is still effective enough to enable highly natural and intelligible speech synthesis.

\noindent \textbf{TIMIT VC (OOD):}
The Phonological Tokenizer outperformed all the baselines across all the metrics. Although it was slightly worse than the Voc-only ($\alpha=1$) in F0 corr., it achieved the best performance in all other metrics.
This demonstrates its ability to retain prosodic information while removing source speaker identity (as also shown in Sec. \ref{subsec:discriminative}) and to enable high-quality speech synthesis. 

\noindent \textbf{Expresso VC (OOD):}
Our model outperformed all the baselines in F0 corr. and UTMOS, but SpkSim and WER were not as good as Discrete WavLM and ASR-only ($\alpha=0$). This is likely because these evaluation metrics tend to favor neutral speech over emotional speech. Indeed, listening to our demo site\footnote{\url{https://ondatk68.github.io/onda-demo/projects/phonological-tokenizer}}, you can find that when using these tokens that focus on linguistic information, the output speech sounds neutral, regardless of the speaking style of the input speech.
In contrast, the Phonological Tokenizer reproduces both the target speaker identity and the speaking style of the input speech. This is particularly interesting considering that both our tokenizer and the vocoder were trained without using any emotional speech.

Overall, the results showed that the Phonological Tokenizer is sufficiently useful for speech synthesis tasks. 
Our tokens showed only slight degradation even compared to acoustic token baseline (WavTokenizer) in ID reconstruction.
Also, 
the overall strong performance of our tokens on OOD VC
indicates that the Phonological Tokenizer successfully disentangles  prosodic and speaker information, which is consistent with the ER and SID results in Sec. \ref{subsec:discriminative}.

\subsection{Evaluation on speechLMs}
\begin{table}[tb]
	\centering
    \vspace*{-2mm}
	\caption{SpeechLM performance: sWUGGY and sBLIMP for lexical and syntactic knowledge; sentiment and speaker consistency for awareness to paralinguistic and non-linguistic aspects; and assesments of the quality of generated speech continuations. }
	\label{tab:speechlm}
    \resizebox{\columnwidth}{!}{
	\begin{tabular}{l@{\hspace{6pt}}l@{\hspace{3pt}}l|c@{\hspace{6pt}}c|cc|cc}
	  \toprule
                & & & \multicolumn{2}{c|}{\multirow{2}{*}{ZeroSpeech}} & \multicolumn{2}{c|}{SALMon} & \multicolumn{2}{c}{\multirow{2}{*}{Speech Continuation}}\\
               &  & & \multicolumn{2}{c|}{} & \multicolumn{2}{c|}{(consistency)} &\multicolumn{2}{c}{} \\
              &   &              & sWUGGY & sBLIMP & Sent.  & Spk & GenPPL & UTMOS \\
              &   & & ($\uparrow$) &  ($\uparrow$) & ($\uparrow$) & ($\uparrow$) & ($\downarrow$) & ($\uparrow$) \\\midrule
	\multirow{3}{*}{Baseline} 
         & Discrete WavLM & (phonetic) & 68.6& 57.1& \textbf{80.5}& \textbf{86.0} & 5.81& 3.60  \\
           & SpeechTokenizer & (hybrid) &66.4 &54.4 & 59.5 & 65.0 & 5.73& 3.64 \\
           & WavTokenizer & (acoustic) &52.5 & 49.3& 66.0& 74.0&6.34 & 2.57 \\\midrule
        Single-task& ASR-only \cite{onda2025differentiablekmeansfullyoptimizeddiscrete} &($\alpha = 0 $) &\textbf{70.0} & \textbf{59.7}& 61.0& 61.0& \textbf{5.60}& 3.56 \\   
        Optimized & Voc-only  & ($\alpha = 1 $) &56.9 & 51.2& 62.5& 79.5& 6.40& 3.67\\ \midrule
               \multirow{2}{*}{\textbf{Proposed}}    & \textbf{Phonological}  & \multirow{2}{*}{$(\alpha = 0.1$)}  & \multirow{2}{*}{67.0}& \multirow{2}{*}{55.2}& \multirow{2}{*}{67.5}& \multirow{2}{*}{66.0}& \multirow{2}{*}{\textbf{5.60}}&\multirow{2}{*}{\textbf{3.86}}  \\   
               & \textbf{Tokenizer} &&&&&&&\\
                    
	  \bottomrule
	\end{tabular}
    }
    \vspace*{-4mm}
\end{table}
Lastly, we trained speechLMs using the obtained tokens and evaluated their performance. We used the slam recipe \cite{maimon-etal-2025-slamming} based on Qwen2.5-0.5B \cite{qwen2}. We used the 6,000-hour subset of LibriLight \cite{librilight} as training data and trained each model for 4 epochs.
As evaluation metrics, we adopted sWUGGY and sBLIMP from the Zero Resource Speech Challenge \cite{dunbar21_interspeech} to assess lexical and syntactic knowledge. To measure awareness to paralinguistic and non-linguistic information, we used SALMon’s sentimental and speaker consistency metrics \cite{maimon2025salmon}. Additionally, to evaluate the quality of speech continuations generated by the speechLMs, we checked generative perplexity (GenPPL) and UTMOS. 
To calculate GenPPL, we first transcribed the generated speech using Whisper-large-v3, and then calculated perplexity using Llama-3.2-1B \cite{grattafiori2024llama3herdmodels}, following \cite{maimon-etal-2025-slamming}.

The results are shown in Table \ref{tab:speechlm}.
The results of ZeroSpeech, which focuses on linguistic information, and the sentimental and speaker consistency results from SALMon, largely align with the ASR, ER, and SID results reported in Sec. \ref{subsec:discriminative}. In ZeroSpeech, the Phonological Tokenizer performed slightly worse than Discrete WavLM and ASR-only ($\alpha=0$), but still outperformed the other baselines. In SALMon, Discrete WavLM achieved exceptionally high scores contrary to the results of ER and SID, but the Phonological Tokenizer exhibited the second highest sentimental consistency, while its speaker consistency was lower than that of WavTokenizer and the Voc-only ($\alpha=1$). 
For speech continuation, the Phonological Tokenizer  achieved the best performance in both GenPPL and UTMOS. These results highlights that the speechLM constructed with our proposed tokens achieved high naturalness in both spoken content and speech quality in the speech continuation task.

\subsection{Ablation study on the vocoder loss weight}
\label{subsec: ablation}
\begin{figure}[t]
  \centering
  \includegraphics[width=\linewidth]{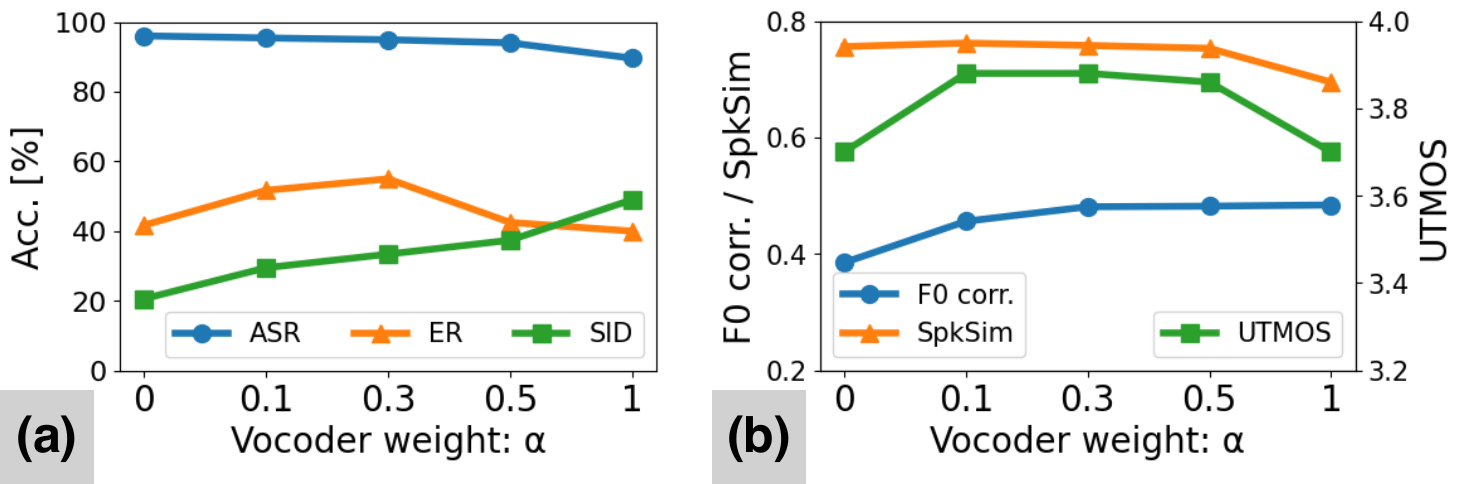}
  \vspace*{-6mm}
  \caption{Ablation results for the vocoder loss weight $\alpha$: (a) Discriminative tasks (ASR, ER, SID), (b) generative task (TIMIT VC).}
  \label{fig:ablation}
  \vspace*{-4mm}
\end{figure}

We conducted an ablation study on the vocoder loss weight $\alpha$ (in Eq.(\ref{eq:joint_loss_voc})). Fig. \ref{fig:ablation} presents the results for the discriminative tasks (ASR, ER, and SID) and the generative tasks (only for TIMIT VC).
For the results of the discriminative tasks, increasing $\alpha$ gradually degrades ASR performance while improving SID. However, ER achieves its optimal value at $\alpha=0.3$. This suggests that increasing $\alpha$ allows both prosodic and speaker information to be encoded in the tokens, but too large value of $\alpha$ make it more difficult to disentangle prosody and speaker timbre. 
For the results of the generative tasks, increasing $\alpha$ leads to higher F0 corr., but when $\alpha=1$, SpkSim decreases largely. The low UTMOS at the two ends can be attributed to different factors: at $\alpha=0$, due to the absence of prosody, and at $\alpha=1$, due to the inclusion of unnecessary speaker information. These results highlight the advantage of our Phonological Tokenizer, which is trained in a multi-task manner to achieve balanced properties between acoustic and phonetic tokens.
\section{Conclusions}
In this work, we propose the \textit{Phonological Tokenizer}, which has intermediate properties between acoustic and phonetic tokens. 
The tokenizer is obtained by fine-tuning phonetic tokens through differentiable k-means using a multi-objective of ASR and speech reconstruction. 
Experimental results across diverse downstream tasks, including speechLMs, demonstrate that the resulting tokens achieve strong performance in capturing linguistic and prosodic information while appropriately disentangling speaker identity. 
Our tokens showed the best results in prosody-sensitive tasks (ER, VC, and speech continuation in speechLM) among all the token tested, and also surpassed hybrid token baseline, SpeechTokenizer, in all the tasks including speech understanding and generation.
Moreover, the fact that our tokenizer relies on a single codebook and trained with only a small amount of data, further underscores the advantages of our method.
Future work includes scaling up the training data for enhanced performance, as well as enabling inference-time controllability for more flexible adjustment of token properties.
\bibliographystyle{IEEEbib}
\bibliography{references}

\end{document}